\documentclass[a4paper,11pt]{article}
\pdfoutput=1 

\usepackage{amssymb,amsthm,booktabs,comment,jheparxiv,float,
stmaryrd,xparse,integrable, graphicx, tikz, tikzscale, tikz-cd}
\notoc

\renewcommand{\d}{\mathrm{d}}
\renewcommand{\u}{\mathrm{u}}

\title{Towards a Carrollian Description of Yang-Mills}

\author{Jeffrey Opreij,}
\author{David Skinner}
\author{\& Hangzhi Wang}

\affiliation{Department of Applied Maths \& Theoretical Physics,\\ University of Cambridge, Wilberforce Road, United Kingdom\vspace{0.1cm}}
\emailAdd{jejjo2@cam.ac.uk}
\emailAdd{d.b.skinner@damtp.cam.ac.uk}
\emailAdd{hw586@cam.ac.uk}

\begin{document}

\abstract{We provide a theory defined purely on null infinity that describes Yang-Mills in the Minkowski space bulk. The dynamical field of our model is the characteristic data of the bulk gauge field, and the action combines an electric branch Carrollian kinetic term with non-local interactions of MHV type that link different points on the celestial sphere. We explicitly show how this theory recovers all MHV and NMHV tree amplitudes in Yang-Mills, and outline how arbitrary N$^k$MHV tree amplitudes may be obtained from its Feynman diagram expansion. The detailed expression we find for the NMHV amplitude appears to be new.}

\maketitle

\newpage


\section{Introduction}
\label{sec:intro}

In the Carrollian approach to holography for asymptotically flat space-times~\cite{Duval:2014uva,Ciambelli:2018wre,Ciambelli:2018ojf,Donnay:2022aba,Donnay:2022wvx,Nguyen:2023vfz,Bagchi:2023fbj,Bagchi:2023cen,Alday:2024yyj,Nguyen:2025zhg,Ruzziconi:2026bix}, one seeks to describe bulk physics using a dual theory defined on null infinity $\scrI$, and often just future null infinity $\scrI^+$. It is natural to expect that such theories should be invariant under the asymptotic symmetry group of the bulk asymptotically flat space-time, viewed as the conformal Carrollian group~\cite{Duval:2014uva,Barnich:2010ojg}. Ward identities associated to this infinite dimensional group impose strong restrictions on correlation functions of any Carrollian theory~\cite{Donnay:2022wvx,Nguyen:2023miw}. In particular, the two-point functions of fundamental Carrollian fields fall into two categories, known as the \emph{magnetic} and \emph{electric} branches, respectively. Fundamentally, these correspond to the ways in which a kinetic term can be constructed using the degenerate metric structure on $\scrI^+$. 

On the magnetic branch, fields are constrained to be constant along generators of $\scrI^+$ (the Bondi $u$ direction), varying only around the celestial sphere. Thus, magnetic branch theories are perhaps best viewed as standard 2d CFTs pulled back to $\scrI^+$ from the celestial sphere. When treated as a theory on $\scrI^+$, integrals over the generators lead to divergences in all correlation functions. While these divergences are easy to regulate, it seems perverse to treat a 2d CFT this way. It may still be possible to recover non-trivial features along the generators of $\scrI^+$ from a magnetic branch dual. Indeed, the Celestial approach to flat holography aims to do exactly this via subtle relations among the conformal dimensions of the celestial operators~\cite{Donnay:2018neh,Donnay:2020guq,Pasterski:2021raf,deGioia:2023cbd}. Still, part of the \emph{raison d'{\^e}tre} of the Carrollian approach is that bulk processes such as a black hole merger, whose emitted gravitational waves impinge on $\scrI^+$ within a certain finite range of $u$ between two long quiescent periods, might be hoped to be more readily described in a dual that lives on $\scrI^+$ itself, rather than just the celestial sphere. From this perspective magnetic branch theories seem rather disappointing.

Conversely, electric branch fields are ultra-local on the celestial sphere and have non-trivial dynamics only along a given generator. This branch has the welcome property of readily providing the $\delta$-functions required when low-point amplitudes describing bulk  scattering of massless particles are written on $\scrI^+$~\cite{Bagchi:2022emh,Mason:2023mti,Nguyen:2023miw}. However, in many regards the electric branch theories seem pathological. Firstly, even in a free electric branch theory, correlation functions of composite operators typically diverge due to products of multiple $\delta$-functions with coincident arguments on the celestial sphere. For the lattice-regulated models and continuum limits studied in~\cite{Cotler:2024xhb}, the resulting theories are generalized free fields, with non-Gaussian correlations suppressed as the regulator is removed. Furthermore, their ultra-local nature prevents an electric branch theory from connecting operators living on different generators. Since massless momentum eigenstates in the bulk limit onto individual generators of $\scrI^+$, it seems impossible for an electric branch Carrollian theory to reproduce any non-trivial bulk scattering amplitude. Perhaps the single-minus amplitudes~\cite{Guevara:2026qzd,Guevara:2026qwa} are an exception in (2,2) signature.

For these reasons, it has often been suggested that any non-trivial (\emph{i.e.} interacting at the quantum level) Carrollian dual of bulk physics must combine both electric and magnetic branches, and may involve some form of non-locality on $\scrI^+$~\cite{Cotler:2025npu}. How to achieve this has remained unclear. 

\medskip

In this paper we report on some partial progress in this direction. We construct a theory living on null infinity that describes Yang-Mills theory (with generic non-Abelian gauge group) in flat Minkowski space. The action combines an electric branch kinetic term with interactions that are non-local across the celestial sphere. Our theory is closely related to the twistor action for Yang-Mills~\cite{Mason:2005zm,Boels:2006ir}, where non-local interactions are familiar: MHV `vertices' are non-local \emph{on twistor space}, but correspond to a perfectly local theory on space-time. In this paper we construct MHV vertices on (good) cuts of $\scrI^+$, which again ensures they give a local space-time theory. As in the twistor string~\cite{Witten:2003nn,Berkovits:2004hg,Seet:2025mes} and the chiral algebra constructions of~\cite{Costello:2022wso,Costello:2022upu}, these MHV vertices can be thought of as the partition function of a 2d chiral CFT that lives on an individual cut and is coupled to the gauge field on $\scrI^+$. Such $u$-independent chiral CFTs can be viewed as a form of magnetic branch theory, though they are more naturally associated to good cuts than to $\scrI^+$.

\medskip

The paper is structured as follows. In section~\ref{sec:geometry} we set notation by reviewing the geometry of $\scrI^+$ that it possesses without reference to a bulk, and the characteristic data of the gauge field from which we build our theory. Section~\ref{sec:action} introduces our action, which lives on a partial complexification $\scrI_\bbC$ where the Bondi $u$ coordinate is allowed to become complex. Our model is not a true Carrollian \emph{dual} in at least two senses. Firstly, the dynamical field is simply the characteristic data of the bulk gauge field. Thus, on-shell excitations of the bulk gauge field are represented by the fundamental fields of our $\scrI$ theory themselves, unlike in AdS where deformations of boundary values of bulk fields source composite operators in the dual. Secondly, the interactions of our theory break supertranslation invariance. Nonetheless, the theory is intrinsically defined on null infinity without reference to a bulk and has the important virtue that it does indeed describe Yang-Mills in bulk Minkowski space-time, at least perturbatively. To test this, in section~\ref{sec:amplitudes} we use our theory to recover Yang-Mills tree amplitudes purely from $\scrI$. We do this explicitly for the $n$-particle MHV and NMHV tree amplitudes, and briefly outline the procedure to compute further N$^k$MHV tree amplitudes. In fact, the detailed form of the NMHV amplitude we find -- given in equation~\eqref{NMHV} -- appears to be new, and we carefully check that it is correct. In section~\ref{sec:twistor_red} we provide a more detailed discussion of the relation of our model to the twistor description of Yang-Mills by showing that our action arises from a gauge reduction down the fibres of the fibration of twistor space over $\scrI_{\bbC}$. We conclude in section~\ref{sec:discussion} by mentioning some open directions.

\section{A Brief Review of the Geometry of $\scrI^+$}
\label{sec:geometry}

We begin by briefly reviewing those rudiments of the geometry of $\scrI^+$ that we shall need later. In the Carrollian framework, $\scrI^+$ is viewed intrinsically as the real 3-manifold
\begin{equation}
\label{scridef}
	\scrI^+ = \text{Tot}(\mathscr{O}(1,\bar 1)_\bbR \to \mathbb{CP}^1)
\end{equation}
where $\mathscr{O}(1,\bar{1})_\bbR=(\mathscr{O}(1)\otimes\overline{\mathscr{O}(1)})_\bbR$. This is probably best understood in coordinates. We let $[\lambda_\alpha]$ denote holomorphic homogeneous coordinates on $\CP^1$ and let $[\bar\lambda_{\dot\alpha}] = [\bar\lambda_0,\bar{\lambda}_1]$ be their (Lorentzian) complex conjugates. We also let $\u$ be a homogeneous coordinate along the fibres of $\scrI^+\to\CP^1$. These homogeneous coordinates are defined modulo the scaling relations
\begin{equation}
\label{scaling}	
	(\u,\lambda_\alpha,\bar\lambda_{\dot\alpha}) \sim (|b|^2\u,b\lambda_\alpha,\bar{b}\bar\lambda_{\dot\alpha})
\end{equation}
for any $b\in\bbC^*$, where the scaling of $\u$ defines the bundle $\mathscr{O}(1,\bar{1})_\bbR$. Note that this scaling preserves the reality of $\u$. On the coordinate patch  $\{\lambda_0\neq0\}\subset\CP^1$ we can fix this scaling by setting $\lambda_\alpha = (\lambda_0,\lambda_1) \sim (1,\lambda_1/\lambda_0) = (1,z)$, whereupon $\bar\lambda_{\dot\alpha}\sim(1,\bar{z})$. Likewise, the homogeneous fibre coordinate $\u$ can be related to the usual (retarded) Bondi coordinate $u$ once one picks a Bondi frame. That is, we choose a future pointing time-like unit vector $t^{\dot\alpha\alpha}$ and set the Bondi
\begin{equation}
\label{Bondi}
	u = \frac{\u}{\la\lambda|t|\bar\lambda]}
\end{equation}
which is invariant under the scaling~\eqref{scaling}. Note that since $t^{\dot\alpha\alpha}$ is time-like, it can never be orthogonal to the real null vector $\bar\lambda^{\dot\alpha}\lambda^\alpha$. If we make the standard choice $t^\mu = \sigma^\mu_{\dot\alpha\alpha}t^{\dot\alpha\alpha} = (1,0,0,0)$ then $u = \u /(1+|z|^2)$ over the patch $\lambda_0\neq0$.

\medskip

\begin{figure}[t!]
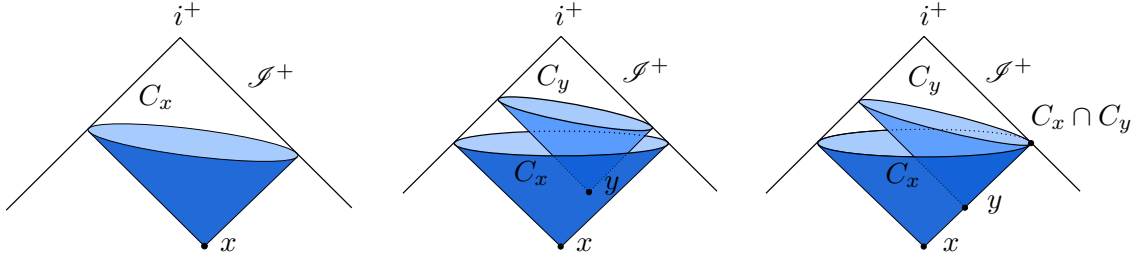

    \centering
    \begin{minipage}{0.33\textwidth}
        \centering
        \includegraphics[height=3.3cm]{scri_2.tikz}

    \end{minipage}\hfill
    \begin{minipage}{0.33\textwidth}
        \centering
        \includegraphics[height=3.3cm]{scri_3.tikz}

    \end{minipage}
    \begin{minipage}{0.33\textwidth}
        \centering
        \includegraphics[height=3.3cm]{scri_1.tikz}

    \end{minipage}
    \caption{(a) \emph{A cut $C_x$ is the intersection of $\scrI^+$ with the future lightcone of a point $x\in\mathbb{R}^{1,3}$.} (b) \emph{Cuts associated to two time-like separated points do not intersect.} (c) \emph{Cuts associated to two null separated points $(x,y)\in\mathbb{R}^{1,3}$  meet on the generator defining the null direction between $x$ and $y$.}}
  \label{fig:scri-triptych}
\end{figure}

In this paper, an important role will be played by \emph{good cuts} of $\scrI^+$. These are the intersections
\begin{equation}
\label{cutdefn}
	C_x = E^+(x)\cap \scrI^+
\end{equation}
of the future light-cone $E^+(x)$ of a bulk point $x\in\bbR^{1,3}$ with $\scrI^+$ and have topology $S^2$. More intrinsically, we can view $C_x$ as a smooth section $\u(\lambda,\bar\lambda)$ which obeys the \emph{good cut} equation~\cite{Newman:1976gc,Tod:2001gcr,Adamo:2009vu}
\begin{equation}
\label{goodcuteqn}
	(D_AD_B u)^{\text{tf}} =0\,,
\end{equation}
where $D_A$ is the Levi-Civita connection for the round metric on $S^2$ and tf denotes the trace-free part. This equation is expressed purely in terms of data on $\scrI^+$. It knows about flat Minkowski space only in the sense that the asymptotic shear, which would in general appear on the \emph{rhs} of~\eqref{goodcuteqn}, is zero. The solutions to ~\eqref{goodcuteqn} are the $\ell=0,1$ spherical harmonics, given in homogeneous coordinates by
\begin{equation}
\label{explicitcut}
	\u = x^{\dot\alpha\alpha}\bar\lambda_{\dot\alpha}\lambda_\alpha\,,
\end{equation}
where, from this perspective, $x^{\dot\alpha\alpha}$ is a constant Hermitian matrix parametrizing the solution.  Given two distinct points $x,y\in\mathbb{R}^{1,3}$ the intersection $C_x\cap C_y$  satisfies
\begin{equation}
\begin{aligned}
	&\text{$(x-y)$ time-like} \quad &\Leftrightarrow \qquad&C_x\cap C_y = \emptyset\,,\\
	&\text{$(x-y)$ null} &\Leftrightarrow \qquad &C_x\cap C_y = \{\text{pt}\}\,,\\
	&\text{$(x-y)$ space-like}  &\Leftrightarrow \qquad &C_x\cap C_y \cong S^1\,.
\end{aligned}
\end{equation}
Generic smooth sections of $\scrI^+\to\CP^1$ play no part in our story, so we shall often abbreviate `good cut' to simply `cut' where this causes no ambiguity.


\medskip

To construct the kinetic term of our action, we will make use of a partial complexification of $\scrI^+$. Following~\cite{Kmec:2024nmu,Kmec:2025ftx}, we define $\scrI_\bbC$ to be the complex 2-fold
\begin{equation}
\label{Cscri}
	\scrI_\bbC = \text{Tot}(\mathscr{O}(1,\bar{1})\to\CP^1)
\end{equation}
where the fibres are copies of $\bbC$ rather than $\bbR$ as on real $\scrI^+$. That is, we allow $\u$ (and the Bondi $u$) to become complex. Note however that we do not complexify the $\CP^1$, so $\bar\lambda_{\dot\alpha}$ still remains the Lorentzian complex conjugate of $\lambda_\alpha$.  Choosing the Bondi frame gives a smooth trivialisation of this bundle by $u = \u/\la\lambda|t|
\bar\lam]$. We equip $\bbC\times\CP^1$ with its product complex structure, in which it has a $\bar\partial$-operator
\begin{equation}
    \bar \partial = \d \bar u \frac{\partial}{\partial\bar u} +  \d\bar{z} \frac{\partial}{\partial\bar z}
\end{equation}
where the angular derivatives act at fixed $u,\bar{u}$.
We also have the holomorphic exterior derivative $\partial = \d u\,\partial_u + \d z\,\partial_z$.

\subsection{Characteristic Data on $\scrI^+$ for Radiative Yang-Mills}

In terms of (retarded) Bondi coordinates for Minkowski space,
\begin{equation}
	\d s^2 = -\d u^2 - 2\d u \d r + r^2 \frac{4 \d z \d\bar{z}}{(1+|z|^2)^2}\,,
\end{equation}
future null infinity $\mathscr{I}^+$ is the surface $r\to\infty$. We assume that the field-strength tensor has the standard peeling behaviour
\begin{equation}
	F_{uz} = \sum_{n=0}^\infty \frac{F_{uz}^{(n)}}{r^n}\,,\quad
	F_{z\bar z} = \sum_{n=0}^\infty \frac{F_{z\bar z}^{(n)}}{r^n}\,,\quad
	F_{ur} = \frac{1}{r^2}\sum_{n=0}^\infty \frac{F_{ur}^{(n)}}{r^n}\,,\quad 
	F_{rz} = \frac{1}{r^2}\sum_{n=0}^\infty \frac{F_{rz}^{(n)}}{r^n}
\end{equation}
as $r\to\infty$, where the $F^{(n)}_{\mu\nu}$ are independent of $r$; \emph{i.e.} they are smooth functions of the coordinates $(u,z,\bar{z})$ on $\scrI^+$. Working in radial gauge $A_r=0$ throughout $\bbR^{1,3}$, this corresponds to the fall-off conditions
\begin{equation}
	A_u = \sum_{n=0}^\infty \frac{A_u^{(n)}}{r^n}\,,\qquad A_z = \sum_{n=0}^\infty\frac{A_z^{(n)}}{r^n}\,,\qquad A_{\bar z}= \sum_{n=0}^\infty \frac{A_{\bar z}^{(n)}}{r^n}\,.
\end{equation}
Imposing $A_r=0$ does not completely fix the gauge; we may still use an $r$-independent (large) gauge transform to set $A^{(0)}_u(u,z,\bar{z})=0$. In this gauge, the pullback of the connection to $\scrI^+$ is
\begin{equation}
	A|_{\scrI^+} = A_z^{(0)} \d z + A_{\bar z}^{(0)}\d\bar{z}\,.
\end{equation}
A Yang–Mills field is called \emph{radiative} if it is a source-free, asymptotically flat solution for which $\scrI^+$ is a good characteristic surface, so that its freely specifiable data live on $\scrI^+$. In practice, this means that the leading sphere connection (or equivalently the functions $A^{(0)}_z(u,z,\bar{z})$ and $A^{(0)}_{\bar z}(u,z,\bar z)$) supplies the radiative characteristic data, and the Yang–Mills equations then determine the rest of the asymptotic expansion recursively once the gauge and asymptotic boundary conditions are fixed (see \emph{e.g.}~\cite{He:2015zea,Nagy:2024jua,Adamo:2021dfg}). The asymptotic boundary conditions also fix the Coulombic integration constants. Since our construction below depends only on the characteristic data itself, we will not need these determined subleading fields explicitly, and we write
\begin{equation}
	A|_{\scrI^+} = a + \bar{a}\,,
\end{equation}
where $a=A_z^{(0)}\d z$ and $\bar{a}=A_{\bar z}^{(0)}\d \bar{z}$.

As a very simple example, consider the case of a momentum eigenstate obeying the linearized Yang-Mills equations on $\bbR^{1,3}$, \emph{i.e.} a massless field of helicity $\pm1$. On $\scrI^+$ these are determined by gauge fields that are localized at a point on $\CP^1$ and have a plane wave dependence in the $u$ direction. Suppose we have a linearized field of helicity $+1$ and null momentum $k_{\dot\alpha\alpha} = \bar\kappa_{\dot\alpha}\kappa_\alpha$. On the patch $\kappa_0\neq0$ set $\kappa_\alpha = \kappa_0(1,z_k)$ with $\omega_k = \la\kappa|t|\bar\kappa]$ the corresponding frequency, then $A|_{\scrI^+}=\bar{a}$ where\footnote{We use the standard conventions that $\bar\delta(z) = \bar\partial (1/z) = \d\bar{z}\,\partial_{\bar z}(1/z)$, with $\partial_{\bar z}(1/z) = \pi\delta^2(z)$ and $d^2z = \frac{i}{2}dz\wedge d\bar{z}$. Likewise, in~\eqref{helicityminus1eigenstate}  $\delta(\bar{z})$ is the (1,0)-form $\d z\,\partial_z (1/\bar{z})$.}
\begin{subequations}
\label{momentumeigenstates}
\begin{equation}
\label{helicity1eigenstate}
	\bar{a} = \frac{1}{\kappa_0^2}\bar\delta(z-z_k)\,  e^{{\rm i}\omega_k u} 
	= \frac{\la \lambda|t|\bar\kappa]}{\la\kappa|t|\bar\kappa]} \,\bar\delta(\la\kappa\lambda\ra) \,\exp\left({\rm i}\frac{\la\kappa|t|\bar\kappa]}{\la\lambda|t|\bar\lambda]}\u\right)
\end{equation}
on $\scrI^+$.  Likewise, a linearized field of the same momentum but helicity $-1$ is given by the characteristic data
\begin{equation}
\label{helicityminus1eigenstate}
	a = \frac{1}{\bar\kappa_0^2}\delta(\bar{z}-\bar{z}_k)\,e^{{\rm i}\omega_k u}
	=\frac{\la \kappa|t|\bar\lambda]}{\la\kappa|t|\bar\kappa]} \,\delta([\bar\kappa\bar\lambda]) \,\exp\left({\rm i}\frac{\la\kappa|t|\bar\kappa]}{\la\lambda|t|\bar\lambda]}\u\right)\,.
\end{equation}
\end{subequations}
For real incoming momenta we use the corresponding negative frequency representatives; equivalently one may write $k=\varepsilon\,|\kappa\rangle[\bar\kappa|$ with $\varepsilon=\pm1$ and include this sign in the frequency.

For linearized fields such as the momentum eigenstates in eqs~\eqref{momentumeigenstates},  the Kirchhoff-d'Adh\'emar formula~\cite{Penrose:1985bww,Mason:1986tn,Adamo:2021dfg} allows us to reconstruct the bulk curvature at a point $x\in\bbR^{1,3}$  from an integral of the characteristic data over $C_x$. Specifically, if we decompose the curvature $F=F_{\dal\al\db\beta}\,\d x^{\dal\al}\wedge\d x^{\db\beta}$ into its self-dual and anti-self-dual parts as
\begin{equation}
    F_{\dal\al\db\beta}(x) =\eps_{\dal \db}  F_{\al \beta}(x) + \eps_{\al\beta} F_{\dal \db}(x) 
\end{equation}
then the Kirchhoff-d'Adh\'emar integrals give
\begin{align}
    F_{\al \beta}(x) &= \int_{C_x}[\bar \lam \d \bar \lam] \,\lam_{\al} \lam_{\beta} \left.\frac{\partial^2 a}{\partial \u^2}\right|_{C_x}\,, \label{ka1} \\
   F_{\dal \db}(x) &= \int_{C_x} \la \lam \d \lam \ra \, \bar \lam_{\dal} \bar \lam_{\db} \left.\frac{\partial^2 \bar{a}}{\partial \u^2}\right|_{C_x}\,, \label{ka2}
\end{align}
where the derivatives are pulled back to the cut by imposing~\eqref{explicitcut}. By differentiating under the integral, it is easily checked that these curvatures obey the linearized equations of motion $\p^{\dal\al} F_{\al\beta}=0$ and $\p^{\dal\al}F_{\dal\db}=0$ for any\footnote{In some circumstances, one may wish to impose fall-off conditions on $A|_{\scrI^+}$ as $|u|\to\infty$; we ignore these throughout this paper.} smooth $a$ and $\bar{a}$. In the particular case of~\eqref{momentumeigenstates}, we recover the standard expressions $F_{\al\beta}= \kappa_\al\kappa_\beta \,e^{{\rm i}k\cdot x}$ and $F_{\dal\db}=\bar\kappa_\dal\bar\kappa_\db\,e^{{\rm i}k\cdot x}$.

\section{A Yang-Mills Action on Null Infinity}
\label{sec:action}

In this section, we construct an action on $\scrI_\bbC$ that corresponds to Yang-Mills theory in the Minkowski space bulk. The action consists of an electric branch kinetic term together with non-local MHV-type interactions and is closely related to the twistor action for Yang-Mills~\cite{Mason:2005zm,Boels:2006ir,Adamo:2011pv}. For a true Carrollian dual in the spirit of AdS/CFT, this action should really involve new fields that live purely on null infinity, which perceive the characteristic data $A|_{\scrI^+}=a+\bar{a}$ as a background gauge field. Instead, as a proxy our action treats the asymptotic data $a,\bar{a}$ themselves as the dynamical fields. To do this, we first extend $a$ to be an arbitrary smooth (1,0)-form on the partially complexified $\scrI_\bbC$, again with $\partial_u \lrcorner a = \partial_{\bar{u}}\lrcorner a=0$ so that $a=a_z(u,\bar{u},z,\bar{z})\d z$.  Similarly, we extend $\bar{a}$ as the (0,1)-form $\bar{a}_{\bar z}(u,\bar{u},z,\bar{z})\d\bar{z}$ on $\scrI_\bbC$. The fields $a$ and $\bar{a}$ are thus partial connections on $\scrI_\bbC$.

\medskip

The action takes the form
\begin{equation}
\label{action}
	S = S_{\text{kin}} + {\rm g}^2 S_{\text{MHV}}
\end{equation}
where ${\rm g}$ is the Yang-Mills coupling. We choose the kinetic part of the action to be
\begin{equation}
\label{Skinetic}
	S_{\text{kin}}[a,\bar{a}] = \int_{\scrI_\bbC} \tr(\partial a\,\bar\partial\bar{a})\,,
\end{equation}
Note that since $a$ and $\bar{a}$ each point only along the $\CP^1$ directions of $\scrI_\bbC$, the derivatives must act in the $\u$ (or $\bar\u$) directions. The kinetic Lagrangian is thus the (2,2)-form
\[
	\tr(\partial a\wedge\bar\partial\bar{a}) = -\d z\wedge\d\bar{z} \wedge\d u\wedge\d\bar{u} \ \tr\!\left(\frac{\partial a_z}{\partial u} \frac{\partial \bar{a}_{\bar z}}{\partial \bar{u}
    }\right).
\]
The equations of motion imply that the fields are harmonic along the fibres of $\scrI_\bbC\to\CP^1$, but remain arbitrary smooth 1-forms in their dependence on $\CP^1$. Such fields restrict to real-analytic data wherever they are regular across the real slice. In particular, the momentum eigenstates~\eqref{momentumeigenstates} used below admit the required analytic continuation. General smooth radiative data instead require a boundary-value formulation, with positive- and negative-frequency parts continued into opposite half-planes; we do not construct that global formulation here.

As in the twistor action for Yang-Mills, our MHV interaction $S_{\text{MHV}}$ is built from \emph{holomorphic frames} on a cut $C_x$. Recall that $\bar{a}$ is the (0,1)-form part of the gauge connection on $\scrI_\bbC$. Noting that $C_x\cong\CP^1$, we define the holomorphic frame\footnote{We work perturbatively around the trivial connection and assume the induced holomorphic bundle on every relevant cut is holomorphically trivial.} $U_x(\lambda,\lambda')$ by the conditions
\begin{subequations}
\begin{align}
	\left.(\bar\partial_{C_x} + \iota^*_x\bar{a})\right|_{C_x} U_x(\lambda,\lambda') &= 0\,, \label{holframe}\\
	U_x(\lambda',\lambda') &=I\,, 
\end{align}
\end{subequations}
where $\bar\partial_{C_x}$ is the Dolbeault operator of the Riemann sphere $C_x\cong\CP^1$ and $\iota_x:C_x\hookrightarrow\scrI_\bbC$ its inclusion into complexified null infinity. The holomorphic frame is analogous to a Wilson line, but for a Riemann surface $\bbC^\times\cong\CP^1\setminus \{\lambda,\lambda'\}$ rather than a real curve. In particular, under a smooth gauge transformation $\bar{a}\mapsto g\bar\partial g^{-1} + g\bar{a}g^{-1}$ on $\scrI^+$, we have
\begin{equation}
	U_x(\lambda,\lambda') \mapsto g(\lambda)U_x(\lambda,\lambda') g^{-1}(\lambda')\,,
\end{equation}
so that the holomorphic frame is a map $U_x(\lambda,\lambda'): E_{(\u',\lambda')} \to E_{(\u,\lambda)}$ between two fibres of a gauge bundle on $\scrI^+$ via holomorphic parallel transport along a cut $C_x$ containing $(\u,\lambda)$ and $(\u',\lambda')$. Solving~\eqref{holframe} perturbatively in $\bar{a}|_{C_x}$ gives
\begin{equation}
\label{holframeexplicit}
	U_x(\lambda,\lambda') = I + \sum_{n\geq 1} \frac{{\rm i}^n}{(2\pi)^n}\int_{C_x^{\otimes n}} \frac{\la\lambda\lambda'\ra\,\la\lambda_1\d\lambda_1\ra \cdots \la\lambda_n\d\lambda_n\ra}{\la\lambda\lambda_1\ra\la\lambda_1\lambda_2\ra\cdots\la\lambda_{n-1}\lambda_n\ra\la\lambda_n\lambda'\ra}\, \bar{a}(\lambda_1)\cdots\bar{a}(\lambda_n) 
\end{equation}
where the $\bar{a}$ here are implicitly pulled back to the cut $C_x$ via~\eqref{explicitcut}. Such holomorphic frames also play an important role in the twistor description of Yang-Mills theory;  see \emph{e.g.}~\cite{Adamo:2011pv,Adamo:2017qyl} for further details.

The MHV interaction is now straightforward to define. We define the real part of the differential operator in the $\u$ direction via
\begin{equation}
     \partial^{\text{R}}_{\u} =\partial_{\u} + \bar \partial_{\u}
\end{equation}
and we set
\begin{equation}
\begin{aligned}
\label{MHVvertex}
    S_{\text{MHV}}[a,\bar{a}] = \frac{1}{2}\int \mathrm{d}^4 x \int_{C_x \times C_x} &\hspace{-0.5cm}[\bar{\lambda} \,\d\bar{\lambda}][\bar{\lambda}' \,\d \bar{\lambda}'] \langle \lambda \lambda' \rangle^2\, \\&\tr\!\left( \left(\partial_{\u}^{\text{R}}\partial_{\u} a(\u,\lambda)\right) U_x(\lambda, \lambda')  \left(\partial_{\u'}^{\text{R}}\partial_{\u'} a(\u',\lambda')\right) U_x(\lambda', \lambda) \right)\,,
\end{aligned}
\end{equation}
where again each $a$ is (implicitly) pulled back to the cut $C_x$ after taking the derivatives. Note that this expression is quadratic in $a$, which appears only through its $\u$ derivatives, and non-polynomial in $\bar{a}$, which appears in the holomorphic frames. The integral $\int \d^4x$ may be thought of intrinsically on null infinity as an integral over the space of solutions to the good cut equation~\eqref{goodcuteqn}. Since our theory naturally lives on the complexification $\scrI_\bbC$, we must specify a four-real-dimensional integration cycle in the space of all complex good cuts of $\scrI_\bbC$ over which to integrate. We make the obvious choice and insist that the cut equation $\u = x^{\da\al}\bar\lambda_\da\lambda_\al$ involves a Hermitian $x^{\da\al}$, corresponding to a point $x$ in real Lorentzian space-time. Such cuts live entirely in the real slice $\scrI^+$ where $\u=\bar\u$. The entire functional ${\rm g}^2S_{\text{MHV}}$, including its quadratic term, is treated perturbatively as an interaction. The propagator below inverts $S_{\text{kin}}$ alone.

\medskip

Since both $S_{\text{kin}}$ and $S_{\text{MHV}}$ involve bare (non-covariant) $u$ and $\bar{u}$ derivatives of the fields, the action is invariant under gauge transformations $g$ only when $g$ is pulled back from a gauge transform on $\CP^1$. That is, the action is invariant under
\begin{align}
    a &\mapsto gag^{-1}  + g\partial g^{-1} \\
    \bar{a} &\mapsto g\bar{a}g^{-1} + g\bar\partial g^{-1}
\end{align}
where $g=g(\lambda,\bar\lambda)$ can be smooth over $\CP^1$, but must be independent of $u$ and $\bar{u}$. This amounts to treating the leading soft modes $J[0,0](z)$ and $\bar{J}[0,0](z)$ of the S-algebra~\cite{Guevara:2021abz,Costello:2022wso} as gauge transformations, while modes $J[m,n](z)$ and $\bar{J}[m,n](z)$ for $m,n\in\mathbb{N}_0$ with $m+n>0$ are retained as non-trivial modes of the fields. We note that dropping the $J[0,0](z)$ is a consistent truncation of the S-algebra.
 
The kinetic part of the action is invariant under the (extended) BMS transformations
\begin{equation}
\label{eBMS}
    \delta u = f(z,\bar{z}) + \frac{u}{2}D_AV^A\,, \qquad \delta z = V^z\,,\qquad \delta\bar{z} = V^{\bar{z}}\,,
\end{equation}
where $f$ is the supertranslation in the chosen Bondi frame; its homogeneous representative $F= \la\lambda|t|\bar\lam] f$ has weight $(1,1)$.  The superrotation $V^A=(V^z,V^{\bar z})$ is a meromorphic vector field on $\CP^1$. In the usual Carrollian nomenclature~\cite{Nguyen:2025zhg,Ruzziconi:2026bix}, the fact that $S_{\text{kin}}$ involves derivatives only in the fibre directions of $\scrI_{\bbC}\to\CP^1$ implies this is a (complexification of) an \emph{electric branch} kinetic term. However, because $S_{\text{MHV}}$ involves integrating over cuts of the form~\eqref{explicitcut}, this term is invariant only under supertranslations $F = \delta c^{\dal\al}\bar\lambda_\dal\lam_\al$ corresponding to bulk translations. Thus, the MHV interaction breaks the extended BMS group down to Poincar{\'e} invariance. We believe this is an important feature of any theory on null infinity that wishes to recover perturbative Yang-Mills amplitudes on Minkowski space: Yang-Mills amplitudes in flat space-time are not the same as those on a generic asymptotically flat background, so any theory of Yang-Mills on $\scrI$ must pick out a preferred background. 

\subsection{The propagator on $\scrI_\bbC$}

\begin{figure}[t!]
    \centering
    \includegraphics[width=0.4\textwidth]{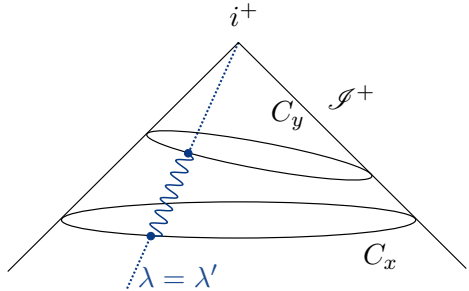}
    \caption{\emph{The electric branch propagator joins points that are on the same generator of $\scrI^+$.}}
  \label{fig:prop}
\end{figure} 

The ultra-local nature of $S_{\text{kin}}$ means that, as far as the kinetic operator is concerned, $a$ and $\bar{a}$ behave just as if they were complex scalars in the complex $u$ plane. Their propagator is thus\footnote{We choose the logarithmic Green function, with additive constant fixed by writing $\ln(|u-u'|^2/\ell_0^2)$. The arbitrary scale $\ell_0$ drops out of the differentiated propagator used to sew vertices.}
\begin{equation}
\label{propagator}
	\la a(u,\bar{u},z,\bar{z})\,\bar{a}(u',\bar{u}',z',\bar{z}')\ra = \Delta(u,z;u',z')=
	\frac{\d z\wedge \d \bar{z}'}{2\pi}\, \delta^2(z-z') \,\ln |u-u'|^2
\end{equation}
in Bondi coordinates. The log term is the standard propagator for a complex scalar in 2d, while the $\delta$-function on the sphere tells us that the fields propagate only along generators of $\scrI_\bbC$. This is as expected for an electric branch Carrollian theory. We can equivalently write
\begin{equation}
\label{homogprop}
	\Delta(\u,\lambda;\u',\lambda') = \frac{\la\lambda\,\d\lambda\ra}{2\pi^2} \frac{\la \lam'|t|\bar\lam]}{\la\lam|t|\bar\lam]} \,\bar{\partial}'\!\left(\frac{1}{\la\lambda\lambda'\ra}\right)\, \ln\left|\frac{\u}{\la\lam|t|\bar\lam]}-\frac{\u'}{\la\lam'|t|\bar\lam']}\right|^2
\end{equation}
in terms of homogeneous coordinates, where $\bar\partial'$ is the Dolbeault operator on the $\CP^1$ associated to $|\lambda'\ra$. 

When used to connect MHV vertices, each end of this propagator will be integrated over the respective cuts. The $\delta$-functions localize one of these $\CP^1\times\CP^1$ integrals to the diagonal $\CP^1$. For this choice of free action, only the mixed $a\bar{a}$ propagator is non-zero.

\section{Bulk Tree Amplitudes from Null Infinity}
\label{sec:amplitudes}

In this section, we test the proposal that the action~\eqref{action} does indeed correspond to Yang-Mills theory in $\bbR^{1,3}$ by using it to compute Yang-Mills tree amplitudes. We begin with the simplest case of MHV trees, before proceeding to an explicit calculation of the NMHV trees in section~\ref{sec:NMHV}. The NMHV case is an important check because these are the first amplitudes that are sensitive to the kinetic term of our action, appearing via the propagator~\eqref{propagator}. The precise expression we obtain for the NMHV tree appears to be new, and we spend some time checking that it is indeed correct. We then outline how the action can be used to generate all N$^k$MHV tree amplitudes. While this is the same recipe as usual MHV (or CSW) diagrams~\cite{Cachazo:2004kj,Boels:2007qn,Adamo:2011pv}, there are several differences. Firstly, here the MHV vertices live on cuts of $\scrI^+$ rather than $\CP^1$s in twistor space. Secondly, our propagator is not the same as the axial gauge twistor propagator, nor the CSW prescription.

\subsection{MHV}
\label{sec:MHV}

\begin{figure}[t!]
    \centering
 	\includegraphics[height=3.3cm]{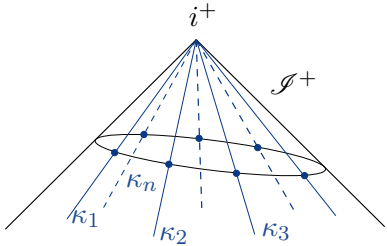}
\caption{\emph{MHV amplitudes come from inserting momentum eigenstates into the MHV vertices in the action and integrating over all good cuts. Note that each momentum eigenstate intersects every cut at a unique point.}}
\label{MHVcutfig}
\end{figure}

The bulk MHV tree amplitudes are essentially trivial to obtain. Expanding out the holomorphic frames using~\eqref{holframeexplicit} and inserting momentum eigenstates~\eqref{momentumeigenstates}, $S_{\text{MHV}}$ is a generating function for the usual Parke--Taylor MHV tree amplitude
\begin{equation}
\label{MHVtree}
	\mathcal{A}_{\text{MHV}}^{(0)}= \la rs\ra^4\,\delta^4\!\left(\sum_{i=1}^n k_i\right)\left[ \frac{\tr(t_1t_2\cdots t_n)}{\la12\ra\la23\ra\cdots\la n1\ra} + \text{non-cyclic perms}\right],
\end{equation}
where gluons $r$ and $s$ have negative helicity. The integrals over each copy of $C_x\cong\CP^1$ are frozen by the $\delta$-functions appearing in the momentum eigenstates, and we are using the standard shorthand $\la ij\ra = \la \kappa_i\kappa_j\ra$.  We remark that the generating function gives the complete amplitude, not just a particular colour-ordering, because the external states may be inserted into the holomorphic frames in any order.

\subsection{NMHV}
\label{sec:NMHV}

Consider an NMHV amplitude in which particles $r$, $s$ and $v$ have negative helicity, with all others having positive helicity. This amplitude will come from a sum of diagrams in which two MHV vertices on cuts $C_x$ and $C_y$ are joined by a propagator. Two of the negative helicity states must be attached to one of the vertices, say $C_x$, with the remaining negative helicity external state on the vertex $C_y$. The propagator then joins the remaining negative helicity field $a$ on the $C_y$ vertex to a positive helicity field coming from the expansion~\eqref{holframeexplicit} of the holomorphic frame on the $C_x$ vertex. 

Without loss of generality, we shall compute the diagram 
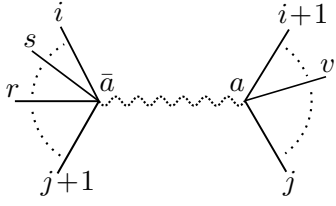
\begin{figure}[h!]
    \centering
    \includegraphics[width=0.3\textwidth]{nmhv.tikz}
\label{NMHV diagram}
\end{figure}

\noindent where external states $L=\{j\!+\!1,\ldots,i\}\supseteq\{r,s\}$ are attached to the vertex $C_x$, while states in the set $R=\{i\!+\!1,\ldots,j\}$ are attached to the vertex $C_y$.  For simplicity we will consider the contribution to the amplitude with colour factor $\tr(t_1t_2\cdots t_n)$. We repeat that the theory generates the full amplitude, summed over all colour orderings.

Since the MHV vertices depend on the negative helicity field $a$ only through $\partial_{\u}^{\text{R}}\partial_{\u}a$, we will need the derivative of the propagator
\begin{equation}
    \la \partial_{u}^{\text{R}}\partial_{u} a(u,z) \, \bar{a}(u',z')\ra = -\frac{\d z \wedge \d\bar{z}'}{2\pi}\,\delta^2(z-z') \left[\frac{1}{(u-u')^2} - \pi \delta^2(u-u') \right]\,.
    \label{eq:MHV_propagator}
\end{equation}
The second term, which is a contact term in the $\u$-direction, comes from the anti-holomorphic derivate in $\u$, and will not contribute to any tree-level scattering amplitudes in the $\rm i\eps$-prescription we will be specifying later. The expressions that follow are to be understood in this prescription, with the regulator temporarily suppressed.

Each end of this propagator is integrated over the respective cut. However, the $\delta$-functions localize the integrals to the diagonal, where both ends of the propagator lie on the same generator of $\scrI^+$ (figure~\ref{fig:prop}). From the expansion~\eqref{holframeexplicit}, the integral over the location of the $\bar{a}(\u',\lambda')$  end of the propagator gives a factor
\begin{equation}
	\int_{C_x}\frac{\la\lambda'\d\lambda'\ra}{\la i\lambda'\ra\la\lambda' j\!+\!1\ra}
	\left.\frac{\partial^2}{\partial\u^2} \Delta(\u,\lambda;\u',\lambda')\right|_{C_x\times C_y} = 
	\frac{\la\lambda\d\lambda\ra}{\la i\lambda\ra\la\lambda j\!+\!1\ra}\frac{1}{\la\lambda|x-y|\bar\lambda]^2}\,.
\end{equation}
This combines with all the external states and remaining terms in the holomorphic frames to give a contribution
\begin{equation}
\label{2MHVvertices}
	\frac{{\rm g}^2}{2\pi}e^{{\rm i}P_L\cdot x}\,e^{{\rm i}P_R\cdot y}\,\text{PT}_n\,\int_{\CP^1}\frac{\la\lambda\d\lambda\ra[\bar\lambda \d\bar\lambda]}{\la\lambda|x-y|\bar\lambda]^2} \,\frac{\la rs\ra^4\la v\lambda\ra^4 \la i\,i\!+\!1\ra\la j\,j\!+\!1\ra}{\la i\lambda\ra\la i\!+\!1\,\lambda\ra\la j \lambda\ra\la j\!+\!1\,\lambda\ra}
\end{equation}
from the diagram joining MHV vertices on the (fixed) cuts $C_x$ and $C_y$. Here we have defined
\begin{align}
P_L = \sum_{\ell\in L} k_\ell\qquad\text{and}\qquad P_R = \sum_{r\in R} k_r
\end{align}
as the sum of external momenta attached to each vertex, and introduced the $n$-particle Parke-Taylor factor
\begin{align}
    \text{PT}_n = \frac{\tr(t_1t_2\cdots t_n)}{\la12\ra\cdots\la n\!-\!1\,n \ra \la n1\ra}\,.
\end{align}
The numerator $\la i\,i\!+\!1\ra\la j\,j\!+\!1\ra$ compensates for the fact that these terms in $\text{PT}_n$ are absent in the MHV diagram we are considering. Two powers of the remaining numerator factor $\la v\lambda\ra^4$ are explicit in~\eqref{MHVvertex}, while the remaining two powers come from the expansion of the holomorphic frames. We can simplify the integrand in~\eqref{2MHVvertices} slightly using the partial fraction identity\footnote{The element $a\in A$ (an external particle adjacent to the propagator in a given colour ordering) is not to be confused with the quantum field $a$. We hope that which is meant is clear from the context.}
\begin{align}
\label{partialfracid}
\frac{\la v\lambda\ra^4 }{\la i\lambda\ra\la i\!+\!1\,\lambda\ra\la j \lambda\ra\la j\!+\!1\, \lambda\ra} = \sum_{a\in A} \frac{\la va\ra^3}{\prod_{b\in A\setminus\{a\}}\la ba\ra}\,\frac{\la v\lambda\ra}{\la a\lambda\ra}\,,
\end{align}
where $A=\{i,i\!+\!1,j,j\!+\!1\}$ is the set of gluons adjacent to the propagator in this diagram. Hence the contribution to the colour-ordered NMHV amplitude from vertices on $C_x$ and $C_y$ may be written as
\begin{equation}
\label{simplifieddiag}
	\frac{{\rm g}^2}{2\pi} e^{{\rm i}P_L\cdot x}\,e^{{\rm i}P_R\cdot y}\,\text{PT}_n\,\la rs\ra^4\sum_{a\in A} \frac{\la va\ra^3\la i\,i\!+\!1\ra\la j\,j\!+\!1\ra}{\prod_{b\in A\setminus\{a\}}\la ba\ra}\int_{\CP^1}\frac{\la\lambda\d\lambda\ra[\bar\lambda \d\bar\lambda]}{\la\lambda|x-y|\bar\lambda]^2} \,\frac{\la v\lambda\ra}{\la a\lambda\ra}\,.
\end{equation}

Let us now consider the remaining integral in~\eqref{simplifieddiag}. If $X=x-y$ is time-like then $\la\lambda|X|\bar\lambda]$ is non-zero for all $|\lambda\ra\in\CP^1$. However, if $X$ is either null or space-like then $\la\lambda|X|\bar\lambda]$ vanishes at points where $C_x$ and $C_y$ intersect. We prescribe the Lorentzian boundary value of the integral over cut separations by replacing\footnote{Here $\sgn(x)$ is the sign function, which we define via
\[
    \sgn(x) = \begin{cases} 1 & x \geq0,\\
                            -1 & x<0\,.
                \end{cases}
\]}
\begin{equation}
\label{Xeps}
X\mapsto X_\epsilon=X-{\rm i}\epsilon t\,\operatorname{sgn}(t\cdot X)\,,
\end{equation} 
where $t$ is the future-pointing time-like vector used before for converting between Bondi $u$ and homogeneous $\u$. With this choice of $\rm i \eps$ prescription, we note the contact term in~\eqref{eq:MHV_propagator} does not contribute to any tree-level amplitude. We then have
\begin{equation}
\label{CP1integral}
\int_{\CP^1}\frac{\la\lambda\d\lambda\ra[\bar\lambda \d\bar\lambda]}{\la\lambda|X_\eps|\bar\lambda]^2} \,\frac{\la v\lambda\ra}{\la a\lambda\ra} = -\frac{2\pi{\rm i}}{X_\eps^2}\frac{\la v|X_\eps|a]}{\la a|X_\eps|a]}
\end{equation}
using the result~\eqref{cp1 int} of the appendix.

With this $\CP^1$ integral in hand, it remains to integrate~\eqref{simplifieddiag} over the locations of the cuts. Let $Y=(x+y)/2$ and $X=x-y$, then the $Y$-integral supplies the momentum-conserving $\delta$-functions, while the remaining integral is a Fourier transform from $X$ to $P_L$. This Fourier transform is not entirely straightforward, and some details are discussed in appendix~\ref{app:Fourier}. The diagram sum must include the bivalent vertex in $S_{\mathrm{MHV}}$, obtained by setting both holomorphic frames to the identity. These two-point vertices require separate treatment: the adjacent labels in~\eqref{partialfracid} are then repeated, so we must cancel the repeated factors in the sphere integrand before evaluating the integral. Including these contributions, the NMHV tree amplitude in the standard colour ordering is
\begin{equation}
\label{NMHV}
\begin{aligned}
\mathcal A_{\text{NMHV}}^{(0)}
={}&
{\rm g}^2\,\delta^4\!\left(\sum_{i=1}^{n}k_i\right)
\mathrm{PT}_n
\Bigg[
{\sum_{\mathrm{diags}}}^{\prime}
\frac{\langle rs\rangle^4}{P_L^2+i\epsilon}
\sum_{a\in A}
\frac{
\langle va\rangle^3
\langle i\,i+1\rangle
\langle j\,j+1\rangle
}{
\displaystyle\prod_{b\in A\setminus\{a\}}\langle ba\rangle
}
\frac{\langle v|P_L|a]}{\langle a|P_L|a]}
\\[2mm]
&\hspace{18mm}
+\sum_{t\in N_-}\langle r_t s_t\rangle^4\frac{
\langle t-1\,t\rangle\langle t\,t+1\rangle[t-1\,t+1]}
{\langle t+1\,t-1\rangle[t\,t-1][t\,t+1]}\Bigg]\,,
\end{aligned}
\end{equation}
where all external labels are understood cyclically modulo $n$. The primed sum includes diagrams with at least two external particles on each vertex. In each such diagram, $L=\{j+1,\ldots,i\}$ contains precisely two negative-helicity particles, labelled $r,s$, while $v$ is the remaining negative-helicity particle; thus these labels are assigned separately in each summand. As before, $P_L=\sum_{l\in L}k_l$ and $A=\{i,i+1,j,j+1\}$. In the second sum, $N_-$ is the set of three negative-helicity external labels, while $r_t,s_t$ are the two elements of $N_-\setminus \{t\}$. Each term is the finite massless limit of the reduced two-pole sphere integral associated with the two-point vertex carrying external leg $t$. Such contributions are necessary in the covariant integral representation, even though two-point vertices make no contribution to the usual CSW representation~\cite{Cachazo:2004kj}. The resulting rational expression contains neither a reference spinor nor the time-like vector used to define the regulator.

\subsubsection{Factorization}\label{factorization}
To the best of our knowledge, \eqref{NMHV} is a new expression for $\mathcal{A}_{\text{NMHV}}^{(0)}$ that has not appeared in the literature before. Note that, although it is built from vertices of MHV type, unlike usual MHV diagrams it does not involve any choice of reference spinor. Nor have we `hidden' the reference spinor by choosing it to coincide with one of the external states: beyond the fact that gluons $r,s,v$ have negative helicity, no  external state is singled out. To check that~\eqref{NMHV} is indeed correct we shall examine its cuts. We first consider the ordinary diagrams in the primed sum in~\eqref{NMHV}. Their propagator-dependent denominators are $P_L^2$ and $\langle a|P_L|a]$ for some external particle $a$. The first class of poles correspond to the expected factorization channels of tree level amplitudes, while the second class must be spurious and cancel in the sum over diagrams. The bivalent terms must also be included when analysing boundary and collinear configurations.

We first check that~\eqref{NMHV} has the expected residue whenever $P_L^2\to0$ for some subset $L=\{j+1,\ldots,i\}$ of the external particles. When $P_L$ is null it can be written as $P_L=|\kappa\ra[\tilde\kappa|$ for some spinors $|\kappa\ra, |\tilde\kappa]$. (The pole may occur in complex momentum space, so $|\tilde\kappa]$ need not necessarily be the Lorentzian conjugate of $|\kappa\ra$.) Dropping the overall momentum conserving $\delta$-function, the residue of~\eqref{NMHV} on such a pole is
\begin{equation}
\begin{aligned}
   &\text{Res}_{P_L^2\to0} \left[\text{PT}_n\,\frac{\la rs\ra^4}{P_L^2}\sum_{a\in A}\frac{\la va\ra^3\la i\,i\!+\!1\ra\la j\,j\!+\!1\ra}{\prod_{b\in A\setminus\{a\}}\la ba\ra}\,\frac{\la v|P_L|a]}{\la a|P_L|a]}\right]\\
   &\qquad=\text{PT}_n\,\la rs\ra^4\,\sum_{a\in A}\frac{\la va\ra^3\la i\,i\!+\!1\ra\la j\,j\!+\!1\ra}{\prod_{b\in A\setminus\{a\}}\la ba\ra}\,\frac{\la v\kappa\ra}{\la a\kappa\ra}
   =\text{PT}_n\,\frac{\la rs\ra^4\la v\kappa\ra^4\la i\,i\!+\!1\ra\la j\,j\!+\!1\ra}{\la i\kappa\ra\la i\!+\!1\,\kappa\ra\la j\kappa\ra\la j\!+\!1\,\kappa\ra}\\
   &\qquad=\la rs\ra^4\, \text{PT}_L \ \la v\kappa\ra^4\, \text{PT}_R\,.
\end{aligned}
\end{equation}
In the second line here we have used the partial fraction identity~\eqref{partialfracid} in reverse to recombine the sum over $A=\{i,i+1,j,j+1\}$. The final line recognizes the result as the product of two separate Parke-Taylor factors associated to MHV diagrams on either side of the cut. Thus the ordinary diagrams have the expected residues in generic multiparticle factorization channels.

\medskip

Next, we prove that the residues of the poles in individual diagrams when $\la i|P_L|i]=0$ for some $i$ cancel in the sum over all diagrams. Consider the diagram where external lines $L_1=\{j+1, \dots, i\}$ are attached to the `left' MHV vertex, and also the diagram with the same colour ordering where $L_2=\{j+1,\ldots,i-1\}$ are attached to the left vertex, so that leg $i$ has been moved to the right. The propagator momenta in the two diagrams are related by $P_{1}=P_{2}+k_i$ so that $\la i|P_{1}|i]=\la i|P_{2}|i]$.

In the case where $i$ is a positive helicity particle, each of these two diagrams has a pole where $k_i\cdot P_{1}=0$. For the first diagram, the residue at the pole is
\begin{equation}
\label{Res1}
    \text{Res}_1=  \frac{\text{PT}_n}{P_1^2}\frac{\la rs\ra^4\la vi \ra^3 \la j\,j\!+\!1\ra}{\la j i \ra \la j\!+\!1\, i \ra} \la v |P_1|i]\,,
\end{equation}
while  the second diagram has the residue
\begin{equation}
    \text{Res}_2 = - \frac{\text{PT}_n}{P_2^2}\frac{\la rs\ra^4\la vi \ra^3 \la j\,j\!+\!1\ra}{\la j i \ra \la j\!+\!1\, i \ra} \la v |P_1|i]\,,
\end{equation}
where we note that $\la v |P_{2}|i] = \la v |P_{1} | i]$ for any $\la v|$. Therefore, in the sum of these two diagrams the coefficient of  $1/\la i|P_{1}|i]$ is
\begin{equation}
    D_1 + D_2 \propto \frac{1}{P_1^2} - \frac{1}{P_2^2} = -\frac{2k_i\cdot P_1}{P_1^2P_2^2} \propto \frac{\langle  i | P_1|i]}{P_1^2P_2^2}
\end{equation}
which cancels the pole.

Now suppose that one of the negative helicity particles $r,s,v$ is adjacent to the propagator. Without loss, we consider the case $\{r,s\}\subset L_1$ but $s\notin L_2$. The residue of the first diagram at $k_s\cdot P_{1}=0$ is again~\eqref{Res1} with $|i\ra$ replaced by $|s\ra$, but the residue of the second diagram now becomes
\begin{equation}
    \text{Res}'_2= - \frac{\text{PT}_n}{P_2^2}\frac{\la sv\ra^4\la rs \ra^3 \la j\,j\!+\!1\ra}{\la j s \ra \la j\!+\!1\, s \ra} \la r |P_{1}|s].
\end{equation}
Writing $\la sv \ra\la r |P_1|s] = \la rv \ra \la s|P_1 |s] + \la sr \ra \la v|P_1 |s]$ the first term in $\text{Res}'_2$ has no pole when $\la s |P_1 | s] = 0$, and again the remaining term cancels the pole in the other diagram.

\medskip

To identify the full amplitude, including the boundary configurations, we use the covariant-integral-to-CSW decomposition of~\cite{Cachazo:2004kj}, whose residue cancellation includes divalent vertices. Together with the evaluations above, this identifies the completed expression with the Yang–Mills NMHV tree amplitude.

\subsubsection{Axial Gauge from a Singular Bondi Frame}
\label{sec:axial}

Since our expression for $\mathcal{A}_{\text{NMHV}}^{(0)}$  is so similar to the standard expression from MHV diagrams
that depends on the CSW reference spinor, it is perhaps worth asking whether we could have obtained exactly this standard form from the theory on $\scrI$. As a slight detour, let us show how to do this.

The usual relation between the Bondi coordinate $u$ and the intrinsic, homogeneous coordinate $\u$ along the fibres of $\mathscr{O}(1,\bar{1})_\bbR\to\CP^1$ is $u = \u / \la \lam |t | \bar \lam]$, where $t$ is a future-pointing time-like vector defining the Bondi frame. Instead, we can define a singular Bondi frame using a null vector $q = |*\ra [*|$. This gives a new Bondi coordinate
\begin{equation}
\label{singularBondi}    
    u_* = \frac{\u}{\la\lam |q|\bar\lam]}
\end{equation}
valid along generators of $\scrI^+$ \emph{except} $|\lam\ra=|*\ra$ where the frame degenerates. The induced metric on each celestial sphere (minus a point) becomes the flat metric.

We now adopt a null-reference sewing prescription with $q=|\ast\rangle[\ast|$, given by~\eqref{nullprescrip}. The singular Bondi coordinate \(u_\ast\) motivates this choice, but the prescription is not obtained by substituting into a pointwise time-ordered kernel on null infinity.
Upon pulling back each end of the propagator to a cut, this amounts to the replacement
\begin{equation}
\label{nullprescrip}
   X\to X_\eps=X- {\rm i} \eps \ \text{sgn}(q\cdot X) \,q
\end{equation}
and we compute the contribution to the amplitude using this new Bondi frame.

In this case, it turns out to be efficient to exchange the order of integrals, computing the Fourier transform before carrying out the $\CP^1$ integral. We have
\begin{equation}
    \frac{1}{(2\pi)^4}\int_{\bbR^{1,3}}\d^4X \,\frac{e^{{\rm i}P_L\cdot X}}{\la\lam|X_\eps|\bar\lam]^2}= 
    \frac{1}{{\rm i}\pi}\,\frac{\la *|P_L| *]^2}{\la\lam *\ra[ *\bar\lam]}\,\frac{\delta^2(\la\lam|P_L|*])}{P_L^2+{\rm i}\eps}\,,
\end{equation}
up to terms that vanish as $\eps\to0^+$. The $\delta$-functions in this expression localize the remaining $\CP^1$ integral. Doing this without using the partial fraction identity gives
\begin{equation}
\begin{aligned}
    &\frac{\la *|P_L|*]^2}{P_L^2+{\rm i}\eps}\int_{\CP^1} \frac{\la\lam\d\lam\ra[\bar\lam\d\bar\lam]}{\la\lam *\ra[*\bar\lam]}\,\frac{\la i\,i\!+\!1\ra\la j\,j\!+\!1\ra\la rs\ra^4\la v\lam\ra^4}{\la i\lam\ra\la i\!+\!1\,\lam\ra\la j\lam\ra\la j\!+\!1\,\lam\ra}\,\delta^2(\la\lam|P_L|*])\\
    =&\frac{1}{P_L^2+{\rm i}\eps}\,\frac{\la i\,i\!+\!1\ra\la j\,j\!+\!1\ra\la rs\ra^4\la v\lam_*\ra^4}{\la i\lam_*\ra\la i\!+\!1\,\lam_*\ra\la j\lam_*\ra\la j\!+\!1\,\lam_*\ra}
\end{aligned}
\end{equation}
where $|\lam_*\ra=P_L|*]$ is exactly the CSW prescription~\cite{Cachazo:2004kj} for the spinor associated to the off-shell propagator. Combined with the overall factor of PT$_n$, this gives the usual CSW representation, or the Feynman diagrams of the twistor action in the axial gauge specified by $|*]$~\cite{Boels:2007qn,Adamo:2011pv}, in which the two-point vertices have vanishing contributions.

\subsection{General Tree Amplitudes}
\label{sec:generaltrees}

The procedure outlined for computing NMHV amplitudes can be straightforwardly extended to more general amplitudes. A connected diagram containing $V$ vertices and $P$ propagators will contribute to the $L$-loop N$^k$MHV amplitude, where $L = P-V+1$ and $k=V-L-1$. This is the standard relation for CSW / MHV diagrams~\cite{Cachazo:2004kj,Boels:2007qn,Adamo:2011pv}, except that here all ingredients are intrinsically defined on null infinity. We outline how to do this in the case of tree amplitudes:
\begin{enumerate}
    \item[\emph{i})] For a tree in the N$^k$MHV sector, set $V=k+1$ and choose cuts $C_{x_v}$, $v=1,\ldots V$. Assign an MHV vertex to each cut and join these vertices using propagators to form a tree. Let $|\lam_{uv}\ra\in\CP^1$ denote the generator of $\scrI^+$ at which the electric branch propagator joining vertices $u$ and $v$ is attached (at both ends). 
    
    \item[\emph{ii})] Distribute the external states among the MHV vertices in some colour ordering. For each MHV vertex, write down a Parke-Taylor factor
    \begin{equation}
        \frac{\la r s\ra^4}{\la a b \ra \cdots \la yz \ra \la za \ra}
    \end{equation}
    where $\{a, \dots z\}$ is the set of all (colour-ordered) lines, both internal and external, connected to the vertex and $r,s$ are the labels of the two negative helicity lines (which may again be an internal line). At each vertex $v$, include a factor of $e^{iK_v\cdot x_v}$, where $K_v$ is the sum of external momenta flowing into $v$.
   
    \item[\emph{iii})]
    Integrate over each $|\lam_{uv}\ra$ using the measure
    \begin{equation}
        \frac{\la \lam_{uv} \d \lam_{uv} \ra [\bar \lam_{uv} \d \bar \lam_{uv} ]}{\la \lam_{uv} |x_u - x_v| \bar \lam_{uv}]^2}\,.
    \end{equation}

    \item[\emph{iv})]
    Integrate over all cut moduli $x_v$, extract the overall momentum-conserving $\delta$-functions and sum over the allowed diagrams, including contributions from bivalent MHV vertices where required.
\end{enumerate}

The sphere integrations require particular care when propagators are adjacent. For sufficiently large $n$, in a generic diagram all propagators will be separated by external states, as in the example
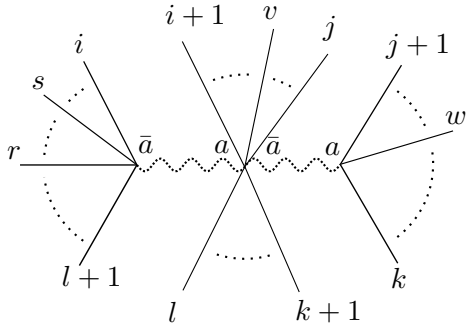
\begin{figure}[H]
    \centering  
    \includegraphics[width=0.4\textwidth]{n2mhv.tikz}
\end{figure}

\noindent at N$^2$MHV. In such diagrams the $\lam$-integrals decouple and may be straightforwardly evaluated as in the previous section. However, the $\lam$-integrals mix in `exceptional' cases where two or more propagators are adjacent, as in the diagram

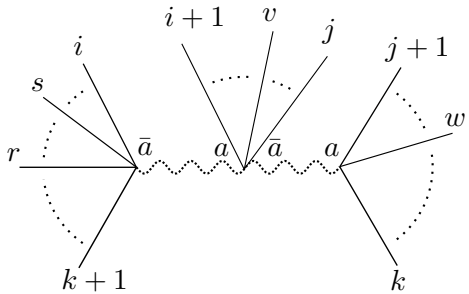
\begin{figure}[H]
        \centering
        \includegraphics[width=0.4\textwidth]{n2mhv_bdry.tikz}\,.
 \end{figure}

\noindent In such cases we encounter integrals of the form
\begin{equation}
    \int \frac{\la \lam \d\lam \ra \la \hat{\lam} \d\hat{\lam} \ra}{\la \lam \hat{\lam} \ra^2 } \frac{\la \lam' \d\lam' \ra \la \hat{\lam}' \d\hat{\lam}' \ra}{\la \lam' \hat{\lam}' \ra^2} \frac{\la v \lam \ra^4 \la w \lam' \ra^4}{\la i \lam \ra \la i\!+\!1\, \lam \ra \la k\!+\!1\, \lam \ra \la \lam \lam' \ra \la j\lam' \ra \la j\!+\!1\, \lam' \ra \la k \lam' \ra}
\end{equation}
with a factor of $\la\lam\lam'\ra$ in the denominator. Performing the integral over $|\lam'\ra$ using~\eqref{cp1 int} leaves us with a sum of integrals for $|\lambda\ra$, some of whose integrands have double poles. Likewise, at N$^k$MHV we may encounter integrals with poles of still higher order. The corresponding integrals can all be done using the result
\begin{equation}
    \int \frac{\la\lam\d\lam\ra\la\hat\lam\d\hat\lam\ra}{\la \lam\hat\lam\ra^2} \left(\frac{\la a \lam \ra}{\la b \lam \ra}\right)^m = -2\pi{\rm i}\left(\frac{\la a \hat b \ra}{\la b \hat b \ra}\right)^m
\end{equation} 
for any $m\in \mathbb{N}_0$, which is shown in~\eqref{higherpoleint}. With this prescription, the sphere integrations give rational expressions in the external spinors and cut-separation data.

\section{Reduction from twistor space}
\label{sec:twistor_red}

In this section we show that the action on $\scrI_\bbC$ we developed in the previous sections can be obtained from a reduction of the twistor action for Yang-Mills~\cite{Mason:2005zm,Boels:2006ir}. The twistor fields are
\begin{equation}
    A\in\Omega^{0,1}(\PT,\mathfrak g) \qquad\text{and}\qquad
    b\in\Omega^{0,1}(\PT,\mathfrak g\otimes\mathscr O(-4))\,,
\end{equation}
and the twistor space action is $S[A,b] = S_1+\rm{g}^2S_2$ where
\begin{equation}
\label{hBFaction}
 S_1[A,b]
 =\int_\PT\mathrm D^3Z\wedge
   \tr\bigl(b\wedge F^{0,2}\bigr)
\end{equation}
with $\mathrm D^3Z$ the weight-four holomorphic volume form on $\PT$ and $F^{0,2}=\bar\partial A + A^2$, while the MHV term is
\begin{equation}
\label{twistor-mhv}
S_2=\frac12\int\limits_{\mathbb R^{1,3}}\d^4x
       \int\limits_{L_x\times L_x}\la\lam\,\d\lam\ra
       \la\lam'\,\d\lam'\ra\la\lam\lam'\ra^2\ 
       \tr\bigl(b(x,\lambda)U_x(\lambda,\lambda')
       b(x,\lambda')U_x(\lambda',\lambda)\bigr)\,.
\end{equation}
in terms the holomorphic frame $U_x$ as in~\eqref{holframe}. Notice that, for Yang-Mills in real Minkowski space, we integrate the MHV term over twistor lines $L_x$ that are real Lorentzian. (See \emph{e.g.}~\cite{Adamo:2013cra} for a review of properties of twistor space and this twistor action.)

\medskip

For the present purposes, it will be important to note that twistor space $\PT$ admits a (non-holomorphic) fibration
~\cite{Kmec:2025ftx}
\begin{equation}
\label{twistorfibration}
  \begin{tikzcd}
    \bbC \arrow{r} & \PT \arrow{d}{p} \\
     & \scrI_\bbC
  \end{tikzcd}
\end{equation} 
over $\scrI_\bbC$, where the projection is given by
\begin{equation}
    p: (\mu^\da,\lambda_\al) \mapsto (\u = [\mu \bar \lam], \lam_\al)\,,
\end{equation}
where $\bar\lambda^\da$ is the Lorentzian conjugate of $\lambda_\al$. Note that fixing $\u$ leaves the freedom $\mu^\da \mapsto \mu^\da+ q \bar \lam^\da$, where  $q$ parametrizes the fibres of this projection, and we can decompose
\begin{equation}
 \mu^\da =\u\,\frac{t^{\da\al}\lam_\al}{\la\lam|t|\bar\lam]} +q\bar\lam^\da \,.
\label{splitting}
\end{equation}
Since $\la\lam|t|\bar\lam]$ never vanishes, this gives a regular smooth splitting over the whole celestial sphere. Under $\lambda\mapsto c\lambda$, we have $\u\mapsto |c|^2\u$ and $q\mapsto(c/\bar c)q$ so that in particular $|q|^2$ is projectively well defined.

It is helpful to introduce an antiholomorphic frame of $(0,1)$-forms adapted to both the fibration $\PT\to\scrI_\bbC$ and the real codimension-1 hypersurface $\mathbb{PN}$ of null twistors lying above real (future) null infinity $\scrI^+\subset\scrI_\bbC$. To choose this antiholomorphic frame, let 
\[
    \rho=\frac{\bar \u-\u}{\la\lam|t|\bar\lam]}\,.
\]
Real null twistors are then the hypersurface
\begin{equation}
    \PN=p^{-1}(\scrI^+)=\{\rho=0\}\subset\PT
\end{equation} 
and we choose the (0,1)-form 
\begin{subequations}
\label{coframe}
\begin{equation}
\label{coframe1}
\begin{aligned}
  \bar e^{\,\u} &= \la\lam|t|\bar\lam]\,\bar\partial_{\PT}\rho \\
 &= \d\bar \u - \bar\u\, \d\log\la\lam|t|\ba\lam]-\bar q\,\la\lam\,\d\lam\ra-q\,[\bar\lam\,\d\bar\lam]\,,
\end{aligned}
\end{equation}
which is the (0,1)-form conormal to $\PN$ inside $\PT$. The complete antiholomorphic frame is then $\{\bar{e}^{\,\u}, \bar e^{\,q}, [\bar\lam\,\d\bar\lam]\}$, where 
\begin{equation}
\label{coframe2}
    \bar{e}^{\, q} = \d\bar q+\bar q\,
     \frac{\la\d\lam|t|\bar\lam]-\la\lam|t|\d\bar\lam]}{\la\lam|t|\bar\lam]}
\end{equation}
\end{subequations}
points along the fibres of $\PT\to\scrI_\bbC$ and $[\bar\lam\,\d\bar\lam]$ is the usual weighted antiholomorphic frame on $\CP^1$. The connection terms here account for the changing spinor basis in
\eqref{splitting} and ensure the correct projective weights.
Despite the appearance of $\d\lam$ in \eqref{coframe}, 
$\bar{e}^{\, \u}$ and $\bar{e}^{\, q}$ are indeed (0,1)-forms on twistor space, because $(\u,q,\lam)$ are not
holomorphic twistor coordinates. For example, direct differentiation gives
\[
 \bar\partial_\PT u=q\frac{[\bar\lambda\,\d\bar\lambda]}{\la\lam|t|\bar\lam]}\,,
 \qquad
 \bar\partial_{\PT}\bar u
     =\frac{\bar e^{\,\u}+q[\bar\lam\,\d\bar\lam]}{\la\lam|t|\bar\lam]}
\]
and their difference verifies $\bar e^{\,\u}= \la\lam|t|\bar\lam]\,\bar\partial\rho$. Using this frame, the Dolbeault operator becomes
\begin{equation}
\label{Dolbeault}
 \bar\partial_{\mathbb{PT}}
   =\bar e^{\,q}\partial_{\bar q}
     +\bar e^{\,\mathrm{u}}\partial_{\bar{\mathrm u}}+[\bar\lambda\,\mathrm d\bar\lambda]\,\bar V_\lambda\,,
\end{equation}
where
\begin{equation}
 \bar V_\lambda
   =q(\partial_{\u}+\partial_{\bar\u})
     -\frac{t^{\dot\alpha\alpha}\lam_\al}{\la\lam|t|\bar\lam]}
       \frac{\partial}{\partial\bar\lam^\da}\,,
\end{equation}
with $\bar\u$ and $\bar q$ derivatives each acting at fixed $(\lam,\bar\lam)$.  Both $\partial_{\bar q}$ and $\bar V_\lambda$ annihilate $\rho$ (they span the CR tangent bundle to $\PN$); this is the reason for the real derivative $\partial_\u+\partial_{\bar\u}$ in $\bar{V}_\lam$.

One reason this frame is useful is because real Lorentzian twistor lines $L_x$ are contained in $\PN$. Since the inclusion $L_x\hookrightarrow \PT$ is holomorphic, we have
\[
 \bar e^{\,\mathrm{u}}|_{L_x} \propto\bar\partial_{L_x}(\rho|_{L_x})=0.
\]
Thus the MHV term is independent of the coefficients of
$\bar e^{\,\u}$ in both $A$ and $b$. These coefficients can consequently serve as Lagrange multipliers in the BF term.

\medskip

We now obtain the action on $\scrI_\bbC$ by reducing the twistor action down this fibration. To carry this out, we must first specify the behaviour at infinity in the non-compact $q$ fibres. Expanding $A=A_{\bar q}\bar e^{\,q}+A_{\bar\u}\bar e^{\,\u}+A_{\bar\lambda}[\bar\lam\,\d\bar\lam]$, and similarly for $b$, we demand
\begin{equation}
 \begin{gathered}
  A_{\bar q} , b_{\bar q}=O(|q|^{-4}),\qquad A_{\bar\lambda},b_{\bar\lambda}=O(1)\\
 A_{\bar{\mathrm u}}=O(|q|^{-1}),\qquad \partial_{\bar q} A_{\bar\u} = O(|q|^{-3}),\qquad b_{\bar{\mathrm u}}=O(|q|^{-2}),
 \end{gathered}
 \label{falloffs}
\end{equation}
as $|q|\to\infty$. We work perturbatively with a fixed boundary trivialization and gauge
transformations preserving these conditions.  Write
$\phi=\langle\lambda\,\mathrm d\lambda\rangle\Phi$, so that $\Phi$ is the coefficient of the horizontal $(1,0)$-form $\phi$. We partially fix the gauge by demanding that the restrictions of $A$ and $b$ to the fibres of $\PT\to\scrI_\bbC$ are both harmonic along the $q$ direction. With the fall-off conditions above, this fixes
\begin{equation}
 A_{\bar q}=0,\qquad
 b_{\bar q}=-\frac{\Phi}{(1+q\bar q)^2},
 \label{fibre-gauge}
\end{equation}
where
\begin{equation}
 \phi=\frac{\langle\lambda\,\mathrm d\lambda\rangle}{2\pi i}
       \int_{C_{\mathrm{u},\lambda}}\mathrm d q\wedge b|_{C_{\mathrm{u},\lambda}}.
 \label{moment}
\end{equation}
The profile in \eqref{fibre-gauge} retains one function on the base $\scrI_\bbC$, whereas the
usual reduction of the twistor action to space-time retains the three components $B_{(\alpha\beta)}$.

Varying $b_{\bar\u}$ now imposes $\partial_{\bar q}A_{\bar\lambda}=0$, whereupon boundedness as $|q|\to\infty$ and Liouville's theorem imply that
$A_{\bar\lambda}$ is independent of $q$. Therefore 
\begin{equation}
\label{a-reduced}
 A=\bar a+\bar e^{\,\u}A_{\bar\u},
\end{equation}
with $\bar a=A_{\bar\lambda}(\u,\bar\u,\lam,\bar\lam)\,[\bar\lam\,\d\bar\lam]$. All remaining fields may vary smoothly over $\scrI_\bbC$; no holomorphy in the complexified Bondi coordinate has been imposed.

The variation of $A_{\bar\u}$ requires more care because its allowed $1/q$ tail can contribute at the fibre boundary. With $\bar\partial_A=\bar\partial+[A,\,\cdot\,]$ and $D_R=\{|q|\leq R\}$,
integration by parts gives
\begin{equation}
 \label{variation}
 \delta_{A_{\bar\u}}S_1
 =\frac{1}{2\pi i}\lim_{R\to\infty}\int\limits_{\scrI_\bbC}\d^2\u
  \la\lam\,\d\lam\ra[\bar\lam\,\d\bar\lam]\ 
    \biggl[\int\limits_{D_R}\d^2q\,\tr\bigl(\delta A_{\bar\u}(\bar\partial_A b)_{\bar q\bar\lam}\bigr) +\oint\limits_{\partial D_R}\d q\,\tr(b_{\bar\lam}\delta A_{\bar\u})\biggr]\,.
\end{equation}
Compactly supported variations first give the bulk equation
$(\bar\partial_A b)_{\bar q\bar\lam}=0$. To solve it, introduce $\chi=\bar q\Phi/(1+q\bar q)$, so that
$\partial_{\bar q}\chi=\Phi/(1+q\bar q)^2$. The solution compatible with the fibre harmonic gauge and boundedness is
\begin{equation}
 \label{compact-variation-sol}
 b\equiv[\bar\lam\,\d\bar\lam]\Psi-\bar\partial_A\chi \pmod{\bar e^{\,\u}}\,,\qquad \psi=\la\lambda\,\d\lam\ra\Psi\,.
\end{equation}
The coefficient $\Psi$, and hence $\psi$, is independent of $q$. Also allowing  the boundary variations in~\eqref{variation} gives
\begin{equation}
 \label{b-boundary}
 \lim_{|q|\to\infty}b_{\bar\lambda}=0\,.
\end{equation}
Although $\chi=O(1/q)$, the $q(\partial_{\u}+\partial_{\bar\u})$ term in
$\bar V_\lam$ has a finite action on it at infinity. The angular and commutator terms decay, leaving
$b_{\bar\lam}\to\Psi-(\partial_{\u}
+\partial_{\bar{\u}})\Phi$.
Hence
\begin{equation}
 \psi=\partial_{\u}^{\mathrm R}\phi\,,\qquad
 (\partial_{\u}^{\mathrm R}\phi)|_{\scrI^+}
       =\frac{\d}{\d\u}\bigl(\phi|_{\scrI^+}\bigr)\,,
\end{equation}
where $\partial_{\u}^{\mathrm R}=\partial_\u+\partial_{\bar\u}$ is an ordinary derivatives along a real generator at fixed celestial
angle. Substitution into the BF term and integration over $q$, gives
\begin{equation}
\label{kinetic}
 S_{\text{kin}}[\phi,\bar a]
   =\int_{\scrI_\bbC}\d\u\wedge\tr
 \bigl(\phi\wedge\bar\partial_{\scrI_\bbC}\bar a\bigr)\,.
\end{equation}
For the MHV term, $p|_{L_x}$ identifies a Lorentzian twistor line with the good cut $C_x$. The normal term in~\eqref{compact-variation-sol} vanishes on $L_x$, and its $\bar\partial_A$-exact term drops out by covariant integration by parts on the compact line, using the holomorphic frame. Only the term proportional to $\psi$ survives. Therefore
\begin{equation}
\label{MHV-reduced}
 S_{\text{MHV}}[\phi,\bar a]
 =\frac12\int_{\bbR^{1,3}}\d^4x
       \int_{C_x\times C_x}
       [\bar\lam\,\d\bar\lam]\,[\bar\lam'\,\d\bar\lam']\,\la\lam\lam'\ra^2\,
\tr\bigl(\partial_{\u}^{\mathrm R}\phi(\u,\lam)
 U_x(\lam,\lam')\partial_{\u'}^{\mathrm R}\phi(\u',\lam')
 U_x(\lam',\lam)\bigr)\,,
\end{equation}
where the fields are restricted to the real cuts after differentiation. The reduced action is $S_{\text{kin}}[\phi,\bar a]+{\rm g}^2S_{\text{MHV}}[\phi,\bar a]$.

To compare with section~\ref{sec:action}, write $\phi=\partial_{\u}a$. The kinetic term~\eqref{kinetic} then becomes~\eqref{Skinetic}, while the interaction becomes precisely~\eqref{MHVvertex}. Therefore, with the fibre conidtions and gauge reductions specified in this section, we have recovered the action~\eqref{action}. We note that the Lorentzian $\rm i \eps$-prescription of \eqref{Xeps} used to sew the MHV vertices together does not immediately arise from this reduction.

\section{Discussion}
\label{sec:discussion}

In this paper we have constructed an action on the partial complexification $\scrI_\bbC$ of $\scrI^+$ and shown that it reproduces Yang-Mills tree amplitudes in $\bbR^{1,3}$. The action consists of an electric branch kinetic term, together with a vertex of MHV type living on (good) cuts.

The action is clearly closely related to the twistor action for Yang-Mills~\cite{Mason:2005zm,Boels:2006ir,Adamo:2017qyl} and we showed how it arises as a gauge reduction of this action. It would be fascinating to see how the relation~\cite{Kmec:2024nmu,Kmec:2025ftx} between S-algebra charge aspects defined in twistor space and in the asymptotic phase space of Yang-Mills~\cite{Freidel:2023gue,Cresto:2025bfo} emerges in our story. It would also be interesting to understand the relation between the action of this paper and the target space string field theory of~\cite{Adamo:2015fwa}. We leave the investigation of these ideas to future work.

The action of this paper, again like the twistor action, breaks manifest parity invariance by including vertices of MHV type only. Indeed, in most of the paper we preferred to write the kinetic term as a functional $S_{\text{kin}}[a,\bar{a}]$ rather than $S_{\text{kin}}[\phi,\bar{a}]$ so as to ensure this term remains parity invariant. It would be interesting to know whether the full $\scrI_\bbC$ action can be written in a manifestly parity invariant way. Certainly, $\scrI_\bbC$ itself can be defined without picking a preferred chirality, unlike twistor space which is inherently chiral. One unwelcome consequence of the fact that our action only contains MHV vertices is that there is currently no way to obtain the 1-loop all plus amplitude of Yang-Mills. In twistor space, this arises from a (non-local) term necessary to cancel anomalous gauge transformations of the chiral twistor action~\cite{Costello:2021bah}. It would be interesting to know whether a similar mechanism can work on $\scrI_\bbC$.

A further point we elided above arises in the continuation from real $\scrI^+$ to $\scrI_\bbC$. It is true that a momentum eigenstate $\bar{a}(u,z)= \bar\delta(z-z_k)\,e^{{\rm i}\omega_ku}$ on $\scrI^+$ can be continued to a solution of the linearized equation of motion $\partial_u\partial_{\bar u} \bar{a}=0$ of $S_{\text{kin}}$ simply by analytic continuation to $\scrI_\bbC$. However, positive-frequency representatives decay as $\operatorname{Im}u\to+\infty$, while negative-frequency representatives decay as $\operatorname{Im}u\to-\infty$.  This is the analogue on $\scrI^+$ of how positive (negative) energy states in $\bbR^{1,3}$ can be continued into the future (past) tube, or of how their twistor states on the Lorentzian slice $\PN\subset\PT'$ can be continued as $H^1$s throughout either $\PT^+$ or $\PT^-$, respectively. At present, we do not have a good understanding of how this fact interplays with our kinetic action as an action on $\scrI_\bbC$. 

\medskip

There are many possible extensions suggested by this work. The most obvious is to construct  an analogous Carrollian description of General Relativity. Here the fundamental data are expected to be the asymptotic shear $\sigma_{AB} = (\sigma_{zz},\sigma_{\bar{z}\bar{z}})=(\sigma,\bar{\sigma})$ encoding the radiative part of the asymptotic connection at $\scrI^+$, or equivalently the trace-free subleading term in the Bondi angular metric.  The shear may appear either directly, or via its $u$ derivatives as the Bondi news, or perhaps via a potential $\chi$ obeying $\sigma = \eth^2\chi$ (see {\it e.g.}~\cite{Winicour:1984,Helfer:2021ktr}). Comparison with the twistor literature~\cite{Mason:2005zm,Mason:2009afn,Sharma:2021pkl} suggests that, in place of the holomorphic frame $U_x(\lambda,\lambda')$ that appeared in the MHV vertex for Yang-Mills, the MHV vertex in gravity should involve integrating insertions of the Bondi news for the negative helicity graviton over a section $Z:\CP^1\to\scrI_\bbC$ that obeys the good cut equation~\cite{Newman:1976gc,Adamo:2009vu} $\bar{\eth}^2Z = \bar{\sigma}(Z,\lambda,\bar\lambda)$ determined by the asymptotic shear due to the positive helicity gravitons. For generic shear, good cuts are complex. A gravitational construction therefore also requires an analytic continuation of the data and a choice of real integration cycle in the resulting solution space~\cite{Newman:1976gc}. A further important extension would  be to understand how to include the possibility of \emph{massive} particles in the bulk scattering process. This would likely shed more light on the boundary conditions one should impose at $\imath^0$ and $\imath^+$.

Finally, we again emphasize that what we have presented is a Carrollian description of Yang-Mills, rather than a true Carrollian dual in the sense of AdS/CFT: the fields from which we have constructed our action on $\scrI$ are the asymptotic values $A|_{\scrI^+}$ of the bulk gauge field, rather than an independent Carrollian CFT. Moving closer to a true Carrollian dual, the MHV vertex on a fixed cut $C_x$ can be viewed as arising from integrating out a theory of chiral free fermions on $C_x$, minimally coupled to $A|_{C_x}$ as a background field. More precisely, as in the original twistor string~\cite{Witten:2003nn}, such a theory leads to our desired MHV vertex only if multi-trace terms are neglected, but the mechanism of~\cite{Seet:2025mes} may be used to cancel such terms automatically. This would be close to a magnetic branch Carrollian theory in the sense that such fermions propagate only along the cut, but to obtain a true magnetic branch one would have to find a way to combine this with the integral $\int\d^4x$ over the space of such cuts. It is unclear to us whether this can be done. The kinetic term seems still more problematic. Although $S_{\text{kin}}$ defines a true electric branch Carrollian theory, it makes the boundary value $A|_{\scrI^+}$ of the gauge field (or rather its extension to $\scrI_\bbC$) a genuine dynamical field in the boundary theory. It may be that this is an inevitable feature of the `leaky boundary conditions'~\cite{Ashtekar:2014zsa,Donnay:2022wvx,McNees:2024iyu} associated with radiative flux in space-times that are asymptotically flat, rather than asymptotically AdS. Nonetheless, we feel it would be preferable if the MHV vertices could be joined using an electric Carrollian theory that was truly independent of the bulk fields. We are reminded of the construction of~\cite{Costello:2019tri}, wherein integrable QFTs are constructed using chiral theories living on 2d defects that talk to one another via a (mixed topological-holomorphic) Chern-Simons theory living in an ambient 4d bulk.

\vspace{1cm}

\noindent {\bf\large{Acknowledgements:}} It is a pleasure to thank Tim Adamo, Lionel Mason, Monica Pate, Paul Luis R{\"o}hl and Romain Ruzziconi for helpful discussions. We also thank the organizers of the 2026 Harvard Workshop on Celestial Holography and the 2026 TAMU Retreat at Cook's Branch, where this work was completed. JO is supported by STFC grant ST/Y509127/1, the Cultuurfondsbeurs (ref 5157191) and  the Fundatie van Renswoude (ref AV20230040). DS is supported by the Simons Collaboration on Celestial Holography and by STFC (UK) grant ST/X000664/1. HW is supported by an STFC Studentship.

\newpage

\appendix
\section{Some $\CP^1$ Integrals}
\label{app:CP1integrals}

In this appendix, $\hat{\ }$ denotes a fixed Euclidean spinor conjugation, with $\hat{\hat\lambda}=-\lambda$. We now prove the identity
\begin{equation}\label{cp1 int}
    I_n = \int_{\CP^1} \frac{\la \lam \d \lam \ra \la \hat \lam \d \hat \lam \ra}{\la \lam \hat \lam \ra^2} \frac{\la \lam a \ra^n}{\la \lam b_1 \ra \cdots \la \lam b_n \ra} =- 2\pi {\rm i}\sum_{i=1}^n\frac{\la b_i a \ra^{n-1}}{\prod_{j \neq i} \la b_i b_j \ra} \frac{\la a\hat b_i \ra}{\la b_i \hat b_i \ra}
\end{equation}
assuming $\la b_i b_j \ra \neq 0$ for $i \neq j$. We first simplify the integrand by decomposing
\begin{equation}
    \frac{\la \lam a \ra^n}{\la \lam b_1 \ra \cdots \la \lam b_n \ra} = \sum_{i=1}^n \lim_{\la \lam b_i\ra \to 0} \left[\la \lam b_i \ra \frac{\la \lam a \ra^{n-1}}{\la \lam b_1 \ra \cdots \la \lam b_n \ra}\right]\frac{\la \lam a \ra}{\la \lam b_i \ra} = \sum_{i=1}^n\frac{\la b_i a \ra^{n-1}}{\prod_{j \neq i} \la b_i b_j \ra} \frac{\la \lam a \ra}{\la \lam b_i \ra}\,.
\end{equation}
Note that this expression is a meromorphic function of $|\lam\ra\in\CP^1$, with only simple poles. The expressions on the left and right have the same poles and the same residues, so by Liouville's theorem they can differ only by a constant. Since both expressions vanish when $|\lam\ra=|a\ra$ this constant must be zero.

Using this identity, the integral reduces to a sum of integrals of the type
\begin{equation}
    I_1 = \int_{\CP^1} \frac{\la \lam \d \lam \ra \la \hat \lam \d \hat \lam \ra}{\la \lam \hat \lam \ra^2} \frac{\la \lam a \ra}{\la \lam b \ra}\,.
\end{equation}
Expanding $|a\ra$ in the basis $\{|b\ra, |\hat{b}\ra\}$ gives
\begin{equation}
\label{intbasis}
    \frac{\la\lam a\ra}{\la\lam b\ra} =  \frac{\la\lam\hat b\ra}{\la \lam b \ra}\frac{\la b a \ra}{\la b \hat b \ra}+\frac{\la a\hat b \ra}{\la b \hat b\ra} \,.
\end{equation}
On the patch $\{\la\lam b\ra\neq0\}\subset\CP^1$ we are free to pick a local coordinate $z = \la\lam \hat{b}\ra/\la\lam b\ra$. With this coordinate $I_1$ becomes
\begin{equation}
    I_1 = \int_{\CP^1} \frac{\d z \wedge \d \bar z}{(1+|z|^2)^2} \left[z\frac{\la b a\ra}{\la b \hat b\ra} + \frac{\la a\hat b \ra}{\la b \hat b \ra}\right] = -2\pi {\rm i} \,\frac{\la a\hat b \ra}{\la b \hat b \ra}
\end{equation}
where the term linear in $z$ vanishes upon integrating over the phase. This shows that
\begin{equation}
    I_n = -2\pi{\rm i}\sum_{i=1}^n\frac{\la b_i a \ra^{n-1}}{\prod_{j \neq i} \la b_i b_j \ra} \frac{\la a \hat b_i \ra}{\la b_i \hat b_i \ra}\,.
\end{equation}
For higher N$^k$MHV tree amplitudes, we also meet integrals of the type
\begin{equation}
    I_{1,m}= \int_{\CP^1} \frac{\la \lam \d \lam \ra \la \hat \lam \d \hat \lam \ra}{\la \lam \hat \lam \ra^2} \left(\frac{\la \lam a \ra}{\la \lam b \ra}\right)^m\qquad\text{for some $m\in\mathbb{N}$.}
\end{equation}
This can be evaluated similarly: taking the $m^{\text{th}}$ power of each side of~\eqref{intbasis} and writing $I_{1,m}$ in the same coordinate patch, all terms with positive powers of $z$ vanish upon integrating over the phase. More precisely, since this integral is not absolutely convergent when $m>1$, we regularize by cutting it off at $|z|=R$ for some $R\gg1$. Terms involving positive powers of $z$ always vanish in this regularized integral. Taking the limit $R\to\infty$ we are left with the circular finite-part of $I_{1,m}$ as
\begin{equation}
\label{higherpoleint}
    I_{1,m} = -2\pi{\rm i}\left(\frac{\la a\hat b\ra}{\la b\hat b\ra}\right)^m\,.
\end{equation}

\section{Fourier Transform of the NMHV Form Factor}
\label{app:Fourier}

The goal of this appendix is to derive that the Fourier transform of
\begin{equation}
{\sum_{\text{diags}}}^{\prime}\left(\frac{\la rs\ra^4}{X_{\eps}^2} \sum_{a\in A} \frac{\la va\ra^3\la i\,i\!+\!1\ra\la j\,j\!+\!1\ra}{\prod_{b\in A\setminus\{a\}}\la ba\ra} \,\frac{\la v|X_{\eps}|a]}{\la a|X_{\eps}|a]}\right)
\end{equation}
is proportional to
\begin{equation}
{\sum_{\text{diags}}}^{\prime}\left(\frac{\la rs\ra^4}{P_L^2+{\rm i}\epsilon} \sum_{a\in A} \frac{\la va\ra^3\la i\,i\!+\!1\ra\la j\,j\!+\!1\ra}{\prod_{b\in A\setminus\{a\}}\la ba\ra} \,\frac{\la v|P_L|a]}{\la a|P_L|a]}\right).
\end{equation}
where $X_\eps = X- {\rm i}\eps t \,\sgn(t\cdot X)$ is the regularized version of the separation $X=x-y$ between two cuts of $\scrI^+$. We will approach this by first showing that distributionally, as $\eps \to 0^+$
\begin{equation}
    \frac{1}{(2\pi)^4}\int \d^4 X  \frac{e^{{\rm i} P_L\cdot X}}{X_{\eps}^2}\frac{\la v|X_{\eps}'|a]}{\la a|X_{\eps}'|a]} 
    =\frac{{\rm i}}{4\pi^2} \frac{1}{P_L^2 + {\rm i}\eps}\,\frac{\la v|P_L|a]}{\la a|P_L|a]- \rm i\eps}
\end{equation}
where in place of $X_\eps$ we have  
\begin{equation}
     X_\eps' = X- {\rm i}\eps t 
\end{equation}
without the factor of $\sgn(t\cdot X)$. The $1/X_{\eps}^2$ term is unchanged. The true regularization $X_\eps$ leads to additional terms which we show vanish when summed over all diagrams. Had we simply replaced $X$ by $X_\eps'$ in the original $\CP^1$ integral~\eqref{CP1integral}, we would be led to the Wightman propagator in momentum space, \emph{i.e.} we would have obtained only the sum of cuts of the amplitude.

Let $k$ denote the null vector normalised by $k\cdot X=\langle a|X|a]$, so that $\la a|X'_\eps|a]= k\cdot X -{\rm i}\eps k\cdot t$. Up to terms that vanish as $\eps \to 0^+$, we compute
\begin{equation}
\begin{aligned}
    \int \d^4 X \frac{e^{{\rm i}P_L\cdot X}\, \la v |X|a]}{(X^2-{\rm i}\eps)(k\cdot X - {\rm i} \eps k\cdot t)} 
    &= {\rm i} \int_0^{\infty} \d \al\, e^{-\al \eps k\cdot t} \int \d^4 X \frac{e^{{\rm i}(P_L - \al k)\cdot X}\la v |X|a]}{X^2-{\rm i}\eps} \\
    &= -8 \pi^2 {\rm i} \int_0^{\infty} \d \al\frac{e^{-\al \eps k \cdot t}\,\la v|P_L|a]}{(P_L^2 - 2 \al k \cdot P_L + {\rm i}\eps)^2} \\
    &= \frac{4 \pi^2 {\rm i}}{(P_L^2 + {\rm i} \eps)} \frac{\la v| P_L|a]}{P_L\cdot k-\rm i \eps}\,.
\end{aligned}
\end{equation}
In going to the second line here we used the standard expression for the regularized Feynman propagator, and the numerator linear in $X$ is implemented by differentiating the scalar Fourier transform with respect to $P_L$. To reach the third line we do a standard contour integral, where we note that as $\eps \to 0^+$ we can take the exponential in the numerator to $1$. 

Now we compute the correction from going to $X_{\eps}$ to $X'_{\eps}$. As a distribution, we have that
\begin{equation}
    \frac{1}{k \cdot X_{\eps}} - \frac{1}{k \cdot X'_{\eps}} = -2 {\rm i} \pi\, \theta(-t \cdot X)\, \delta(k \cdot X)\,.
\end{equation}
The Fourier transform of this difference  equals
\begin{equation}
\label{Xperpint}
    2 \pi {\rm i} \int \d^2 X_{\perp} \frac{e^{{\rm i} P_{L \perp} \cdot X_{\perp}} \la v|X_{\perp}|a]}{X_{\perp}^2+{\rm i}\eps} \left[ \pi \,\delta (k \cdot P_{L}) - {\rm i}\,\text{PV}\frac{1}{k \cdot P_L} e^{-{\rm i} (t_{\perp} \cdot X_{\perp})k \cdot P_{L}/ {k \cdot t}}\right]
\end{equation}
where $t_{\perp}$ and $X_{\perp}$ are the components of $t$ and $X$ orthogonal to both $k$ and any null $q$ such that $k \cdot q = 1$. 

In the sum over diagrams, the $\delta(k \cdot P_L)$ term yields a term proportional to
\begin{equation}
    \sum_{\text{diags}}\left[\frac{\la rs\ra^4}{P_L^2+{\rm i}\epsilon} \sum_{a\in A} \frac{\la va\ra^3\la i\,i\!+\!1\ra\la j\,j\!+\!1\ra}{\prod_{b\in A\setminus\{a\}}\la ba\ra} \,{\la v|P_L|a]}\ \delta(\la a | P_L|a])\right]
\end{equation}
which we know vanishes, since in the full expression the $\la a | P_L|a]$ poles cancel (\textit{cf.} section~\ref{factorization}). The remaining term in ~\eqref{Xperpint} is proportional to $1/\la a|P_L|a]$ and yields a correction term
\begin{equation}
    \Delta \mathcal A^{(0)}_{\text{NMHV}} \propto  \sum_{\text{diags}}\sum_{a\in A} \left({\la rs\ra^4}  \frac{\la va\ra^3\la i\,i\!+\!1\ra\la j\,j\!+\!1\ra}{\prod_{b\in A\setminus\{a\}}\la ba\ra} \frac{1}{k_a \cdot P_L}\left[\,\frac{\la v|\tilde P_{L,a}|a]} {\tilde P_{L,a}^2 + {\rm i}\eps} \right]\right)
\end{equation}
where 
\begin{equation}
    \tilde P_{L,a} = P_L - \frac{\la a| P_L|a]}{\la a| t|a]} t.
\end{equation}
We can use an analogous argument to the one in section \ref{factorization} to show that terms with a prefactor of $1/\tilde P^2_{L,a}$ cancel pairwise with a term with prefactor $1/\tilde P^2_{L',a}$, where $L'$ is associated to a diagram where we shift particle $a$ from the left MHV vertex to the right (or vice versa). We note the identities
\begin{equation}
\begin{aligned}
    \tilde P^2_{L, a} &= \tilde P^2_{L', a} \\
    \la v|\tilde P_{L,a} |a] &= \la v|\tilde P_{L',a} |a]
\end{aligned}
\end{equation}
so these adjacent diagrams' correction terms differ only by a minus sign and thus cancel. This argument identifies the rational expression at generic external momenta. Its extension to distributions on exceptional momentum configurations requires the stated boundary-value and contact-term prescriptions.

All cancellation statements in this appendix refer to the completed diagram sum, including the two-point configurations. When the moved external leg has negative helicity, the corresponding Schouten identity is also used.

\bibliographystyle{JHEP}
\bibliography{Carrollian}

\end{document}